\documentclass[proof]{pasj00}
\draft

\begin{document}
\SetRunningHead{Author(s) in page-head}{Running Head}
\Received{2009 March 26}
\Accepted{}

\title{Suzaku Observation of the Metallicity Distribution \\
in the Elliptical Galaxy NGC 4636}
\author{Katsuhiro H{\normalsize{AYASHI}}$^1$,\ Yasushi F{\normalsize{UKAZAWA}}$^1$,\ Miyako
T{\normalsize{OZUKA}}$^1$ \\ Sho N{\normalsize{ISHINO}}$^1$,\ Kyoko M{\normalsize{ATSUSHITA}}$^2$,\ Yoh T{\normalsize{AKEI}}$^3$,\
and Keith A. A{\normalsize{RNAUD}}$^{4,5}$\\
\itshape{\small{$^1$Department of Physical Science, Hiroshima University, 1-3-1 Kagamiyama, Higashi-hiroshima, Hiroshima 739-8526}} \\
\itshape{\small{$^2$Department of Physics Tokyo University of Science,
1-3 Kagurazaka Shinjuku-ku, Tokyo 162-8601}} \\ 
\itshape{\small{$^3$Institute of Space and Astronautical Science (ISAS),
Japan Aerospace Exploration Agency (JAXA), }} \\
\itshape{\small{3-1-1 Yoshinodai, Sagamihara, Kanagawa, 229-8510}}\\
\itshape{\small{$^4$CRESST and X-ray Astrophysics Laboratory, NASA Goddard
Flight Space Center, 8800 Greenbelt Rd., Greenbelt, MD 20771, USA}}\\
\itshape{\small{$^5$Astronomy Department, University of Maryland,
    College Park, MD 20742, USA}}}
\date{}
\maketitle

\begin{abstract}
NGC 4636, an X-ray bright elliptical galaxy, was observed for 70 ks
with Suzaku. The low background and good energy resolution of the XIS
enable us to estimate the foreground Galactic emission accurately and
hence measure, for the first time, the O, Mg, Si and Fe abundances out
to a radius of $\sim$28 arcmin ($\simeq$ 140 kpc).  These metal
abundances are as high as $>$1 solar within the central 4' and
decrease by $\sim$50\% towards the outer regions.  Further, the O to
Fe abundance ratio is about 0.60--1.0 solar in all regions
analyzed, indicating that the products of both SNe II and SNe Ia have
mixed and diffused to the outer regions of the galaxy.  The O and Fe
metal mass-to-light-ratios (MLR) of NGC 4636 are 2--3 times larger
than those of NGC 1399 implying that metal distributions in NGC 4636
are less extended than those in NGC 1399, possibly due to
environmental factors, such as frequency of galaxy interaction. We
also found that the MLRs of NGC 4636 at 0.1 $r_{180}$ are $\sim$5
times smaller than those of clusters of galaxies, possibly consistent
with the correlation between temperature and MLR of other spherically
symmetric groups of galaxies. We also confirmed a resonant scattering
signature in the Fe ${\rm_{XV\hspace{-.1em}I\hspace{-.1em}I}}$ line in
the central region, as previously reported using the XMM-Newton RGS.
\end{abstract}

\section{Introduction}

The universe is filled with matter in various physical states. One of
these is the hot interstellar medium (ISM) in elliptical galaxies.
The ISM is gravitationally bound by the mass of stars, dark 
matter and itself, and its temperature is about 0.5--1 keV. Emission
from the ISM is observed in the X-ray band and its luminosity reaches about
10$^{39}$--10$^{42}$ erg s$^{-1}$. By accurate analysis of X-ray spectra, 
we can estimate the temperature and metallicity distributions of the
ISM and reveal the metal enrichment history of the universe.
 
In the past, X-ray imaging detectors, such as Chandra ACIS or 
XMM-Newton EPIC, have observed the ISM with high angular resolution and 
high signal-to-noise ratio (S/N) or wide field of view, and revealed
temperature structures, surface brightness and mass distributions 
(e.g., Buote 2002; Buote et al. 2003; Fukazawa et al. 2006; 
Humphrey et al. 2006).
In addition, observations by XMM-Newton have suggested similar
metal abundance patterns to those in the centers of clusters or groups of galaxies
(e.g., Matsushita et al. 2003; Matsushita et al. 2007b). 
The excellent energy resolution 
of XMM-Newton RGS has been used to precisely measure metal abundances 
(e.g., Xu et al. 2002; Tamura et al. 2003) and show that Si and Fe 
abundances in the central region are similar, while the O abundance 
is in general about half that of Fe. 
To understand metal enrichment processes, measurements of 
O, Ne, or Mg abundances in the outer regions are needed.
To unambiguously constrain the O abundance in elliptical galaxies it
is also necessary to have an accurate estimate of the foreground
Galactic thermal emission.
The best detector to satisfy these conditions 
is the Suzaku XIS which has a better energy resolution and a lower background
than any previous X-ray CCD camera.
The launch of the Suzaku XIS (Koyama et al. 2007) with high energy resolution 
in the lower energy X-ray band allows us to obtain O and Mg abundance
distributions through accurate measurements of their emission lines.
Past studies (e.g., NGC1399 ; Matsushita et al. 2007a, NGC 5044 ; 
Komiyama et al. 2008) have shown that the O to Fe abundance 
ratio, which can be used to estimate the number ratio of Ia to II supernovae, 
is about 0.6--0.8 solar in all regions. However, observations in the outer
regions, where metals ejected from galaxies have remained, are still
poor. 
So far, Suzaku measurements of abundance distribution have been reported for 
three elliptical galaxies.
NGC 5044 and NGC 507 exhibit a steeper radial O abundance profile
than NGC 1399 while there is a large scatter among their O mass-to-light ratios.
(Matsushita et al. 2007a; Komiyama et al. 2009; Sato et al. 2009a).
Studies of these properties in additional galaxies will thus aid
understanding of metal enrichment processes.

In this paper, we report the results of spectral analysis of NGC 4636 
observed by the XIS detector onboard Suzaku (Mitsuda et al. 2007). 
NGC 4636 (z = 0.00313;\ $\simeq$ 17 Mpc, Smith et al. 2000) is an X-ray bright 
elliptical galaxy belonging to the Virgo cluster. Metal-enriched hot
gas is extended over the entire galaxy group (Matsushita et al. 1998).
This galaxy is located at the south end of the cluster,
where the X-ray emission from the Virgo center is negligible, so
we can study the metal enrichment history 
of a giant elliptical galaxy (Matsushita et al. 1997).  
So far, observations by ASCA (Matsushita et al. 1997), 
Chandra (Ohto et al. 2003; Humphrey et al. 2006), 
and XMM-Newton (e.g. Xu et al. 2002) 
have been used to measure metal abundances. 
The temperature of hot gas in this galaxy is 0.6--0.7 keV, which is
lower than the 1.0--1.5 keV of NGC 1399, NGC 507, and NGC 5044.
Therefore, we can also discuss the temperature dependence of metal
properties.
In addition, NGC 4636 is quite isolated in comparison with the above 3
galaxies, and thus the comparison will aid investigations of environmental effects.
We describe the observation and data reduction in section 2.
Section 3 on the analysis and results describes the methods we used to 
estimate background and fit spectra as well as the resulting
temperature and metallicity distributions.
Additionally, we discuss the confirmation of
resonance scattering in the Fe
${\rm_{XV\hspace{-.1em}I\hspace{-.1em}I}}$
line in the spectrum of the central region. In section 4,  
we discuss the O, Ne, Mg, Si abundance ratios to Fe, their distributions,
the metal transport process, and resonance
scattering of the Fe 
${\rm_{XV\hspace{-.1em}I\hspace{-.1em}I}}$ line. 
A summary is given in section 5.

Throughout this paper, we use a Hubble constant value 
of $H_{0}$ = 70 km s$^{-1}$Mpc$^{-1}$ or $h_{100}$ = 0.7 
(1 arcmin $\simeq$ 5 kpc),
and a virial radius of
{\it r}$_{180}$ = 1.95$h^{-1}_{100}$ 
$\sqrt[]{\mathstrut k{\langle T\rangle}/10{\rm keV}}$ 
Mpc (Markevitch et al. 1998) where $kT$ is the gas temperature.
The Galactic hydrogen column density is $N_{\rm H}$ = 1.8$\times$10$^{20}$
cm$^{-2}$ in the direction of NGC 4636 (Stark et al. 1992). 
We use the Anders and Grevesse (1989) definition of Solar abundance
ratios for easier comparison with other Suzaku papers which use this definition.
Other recent papers often use instead Grevesse and
Sauval (1998). The former has 26\% and 48\% higher abundances for O
and Fe, respectively. All errors are given as a 90\% confidence range.

\section{Observation and Data Reduction}

A pointing observation of NGC 4636 by Suzaku was carried out on
December 6--7, 2005 for about 70 ks. In order to measure metal abundances in
the outer region of NGC 4636, a 10' north offset observation was 
carried out on December 7--9, 2007. Another north-west offset observation 
at 3.2$^{\circ}$ away from the NGC 4636 center was also carried out 
on June 17--18, 2007 to
estimate the foreground Galactic emission -- this
region contains no other bright objects. 
All the XIS data were taken with both 3$\times$3 and 5$\times$5
pixel mode. The observation log is summarized in Table 1.

\begin{table}[h]
\caption{Observation log}
\begin{center}
\begin{tabular}{cccc}
\hline\hline
Target & NGC 4636 & $^1$NGC 4636-NORTH & $^2$NGC 4636-GALACTIC\\ \hline
z     & 0.00313  & -              & - \\ 
position\ (Ra,\ Dec) in J2000 & 190.7250,\ 2.7520 & 190.6828,\ 3.0503 
                                            & 192.5138,\ 5.4608\\ 
exposure\ time\ (ks) & 70.3   & 51.1 & 34.9 \\ 
date\ (y-m-d\ h:m) & 05-12-06\ 06:38 & 07-12-07\ 05:01 
                                          & 07-06-17\ 20:11\\ \hline
\multicolumn{4}{l}{$^1$NGC 4636-NORTH\ 
                      : 10' offset observation to the north of NGC 4636}\\
\multicolumn{4}{l}{$^2$NGC 4636-GALACTIC\ 
                      : 3.2$^{\circ}$ offset observation for estimating of the
 foreground Galactic emission}\\
\end{tabular}
\end{center}
\end{table}   

We analyzed the public rev 1.2 data and screened them with cut-off rigidity
(COR) $>$ 8 GV, target elevation angle above 
the earth rim (ELV) $>$ 5$^{\circ}$, 
and elevation angle above the day
earth rim (DYE\_ELV) $>$ 20$^{\circ}$. 
For the 2005 observation, we 
extracted spectra of six annular regions centered on the X-ray peak
(Figure 1(a)). For the north offset observation, 
we analyzed the spectrum of a 8'-radius aperture 
centered on a position at 20' away from the galaxy center (Figure 1(b)).
In extracting the above spectra, we excluded the region 
which is contaminated by  
the Fe calibration source. We used xisrmfgen\ version\ 2006-11-26 and
xissimarfgen\ version\ 2006-10-26 to make response 
matrix files (RMF) and ancillary response files (ARF), respectively. The
source mode of xissimarfgen was set to SKYFITS,
which is appropriate for analyzing extended objects, and the XIS
image file was input. Data from the hard X-ray
detector, HXD, are not treated here because no
signal was detected from NGC 4636; an upper limit of the hard X-ray
emission is $9\times10^{-12}$ erg cm$^{-2}$ s$^{-1}$ (10--50 keV) for
the 3\% background uncertainty (Fukazawa et al. 2009).


\begin{figure}[h]
\begin{center}
\begin{tabular}{cc}
\begin{minipage}{0.5\hsize}
\hspace{0.7cm}(a)
\begin{center}
\rotatebox{90}{\scriptsize{\hspace{1.5cm}{Dec(J2000)}}}
\rotatebox{0}{\resizebox{7.0cm}{!}{\includegraphics{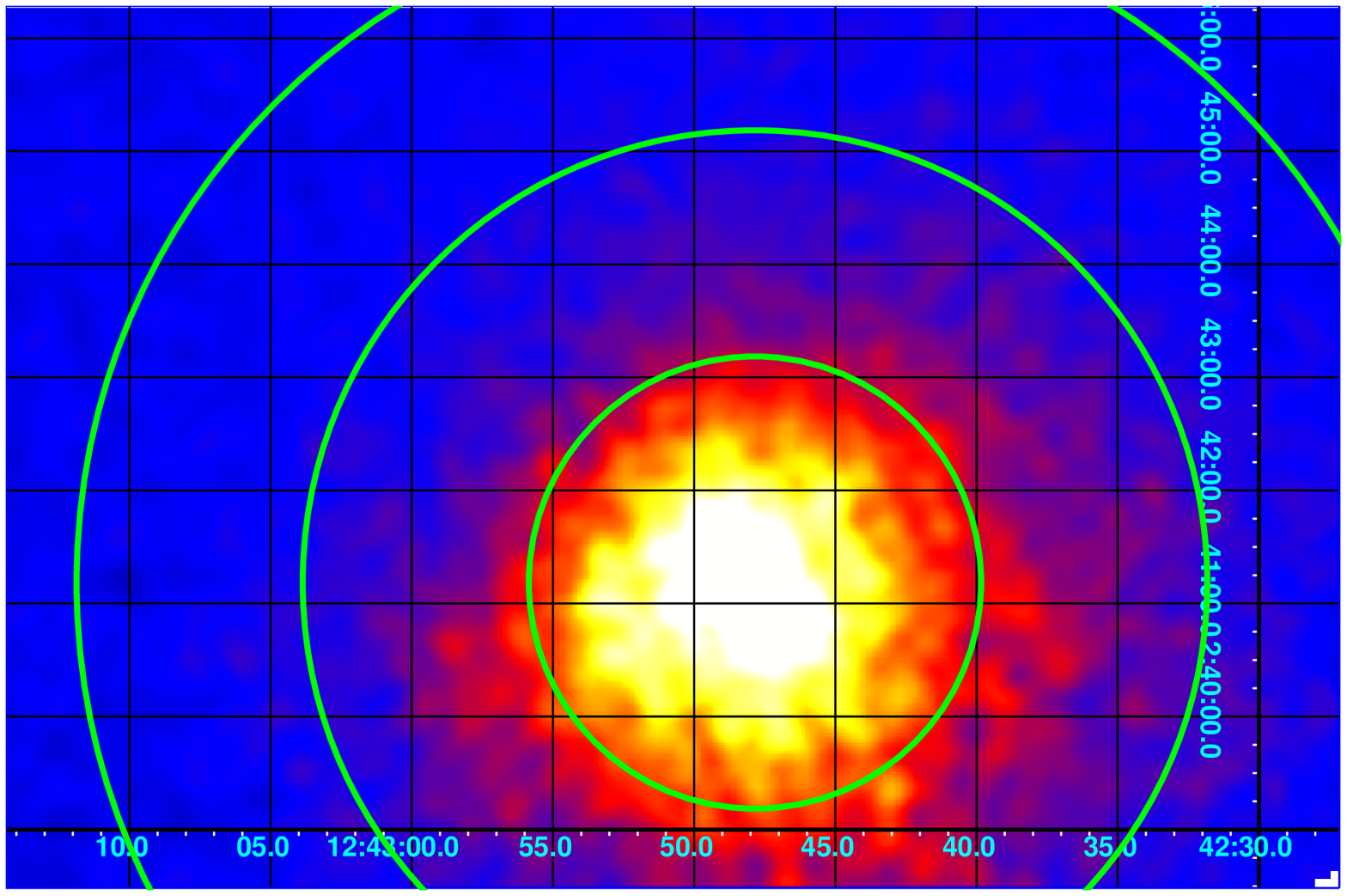}}}\\
{\vspace{-0.3cm}}
{\scriptsize{Ra(J2000)}}
\end{center}
\end{minipage}
\begin{minipage}{0.5\hsize}
\hspace{0.7cm}(b)
\begin{center}
\rotatebox{90}{\scriptsize{\hspace{1.5cm}{Dec(J2000)}}}
\rotatebox{0}{\resizebox{7.0cm}{!}{\includegraphics{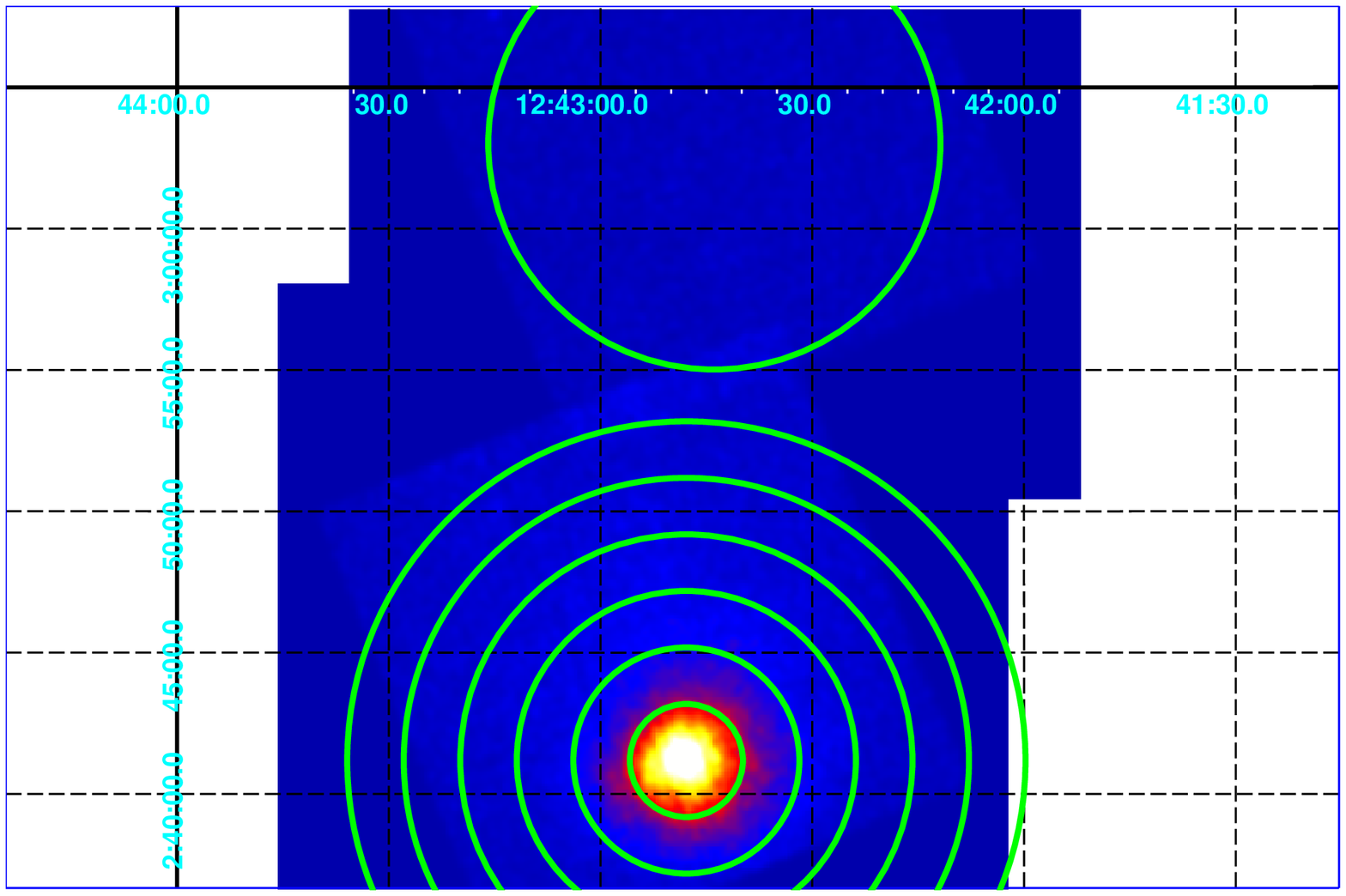}}}\\
{\vspace{-0.3cm}}
{\scriptsize{Ra(J2000)}}
\end{center}
\end{minipage}
\end{tabular}
\end{center}
\caption{X-ray image (0.3--5keV) of NGC 4636, convolved with a
 gaussian of $\sigma =$ 3 pixel. 
The NXB (non X-ray background) is subtracted, 
and the vignetting effects and the difference 
in exposure times are corrected.
(a) Observation of the center region. 
(b) Mosaic image of three observations. Boundaries of analysis regions are shown by a circle. }
\end{figure}


\section{Spectral Analysis and Results}

\subsection{Estimation of Background}

In order to study the emission from NGC 4636 correctly,
we have to estimate the background accurately.
This consists of the instrumental non X-ray
background (NXB), the cosmic X-ray background (CXB) and 
the foreground Galactic X-ray emission (GXE).

First, we estimated the NXB spectra in each analysis region using the tool 
``xisntebgdgen'', which adds COR-sorted 
($<$ 4GV, 4--13 GV in 1 GV step, and $>$ 13 GV) night-Earth spectra 
weighted by the exposure time for each COR in the NGC 4636 observation.
Second, using the sky-averaged CXB parameters from HEAO-1 (Boldt et al. 1987) 
summarized in table 2, we estimated the CXB contribution
by using an uniform-sky ARF, which is generated with
``xissimarfgen'' and assumes that the emission is uniform within 20'.
Finally, we estimated the GXE, using the 
3.2$^{\circ}$ offset observation by extracting spectra within 
8' ($\simeq$ 40 kpc) of the center of the field of view.
We used a uniform-sky ARF as for the CXB estimation.
The GXE consists of a Local Hot Bubble (LHB) component and a Milky Way Halo (MWH)
component. These have temperatures of approximately $\sim$0.08 keV 
and $\sim$0.2--0.3 keV, respectively (e.g., Lumb et al. 2002).
We fitted the GXE spectrum with a two-temperature APEC model, where
temperatures are set to be free and metal abundances are fixed to one solar.
At these temperatures, the continuum emission is vanishingly small
compared to the line emission. 
Therefore, even if this assumption for metal abundance is not completely
correct, the estimation of the GXE will not change significantly.
The results of the fit are shown in table 2, along with the fixed
parameter values used for the CXB model. Figure 2 shows the best fit
model and spectra.
The temperatures obtained are consistent within errors with the
results from the ROSAT-PSPC all-sky survey (Kuntz \& Snowden 2000).
Using the best-fit parameters, we modeled the Galactic emission in 
each spectral region of the NGC 4636 observations by assuming the same
surface brightness and spectral parameters as those of the
3.2$^{\circ}$ offset region.

\begin{table}
\caption{Spectral parameters of the CXB and the GXE} 
\begin{center}
\begin{tabular}{ccccccc} \hline\hline
\multicolumn{7}{l}{Cosmic X-ray Background (CXB)$^{\ast}$}  \\ \hline
\multicolumn{3}{l}{POWER-LAW $\times$ HIGHECUT} 
  & \multicolumn{1}{c}{photon-index} 
  & \multicolumn{2}{c}{normalization$^{\ddagger}$}
  & \multicolumn{1}{c}{e-folding energy (keV)} 
  \\ 
\multicolumn{3}{c}{}
  & 1.29 & \multicolumn{2}{c}{8.21$\times10^{-4}$} & 40.0\\ \hline
\multicolumn{7}{l}{Galactic X-ray Emission (GXE)$^{\dagger}$} \\ \hline
\multicolumn{2}{c}{APEC + APEC} 
  & \multicolumn{2}{l}{Milky Way Hallo (MWH)} 
  & \multicolumn{2}{l}{Local Hot Bubble (LHB)}
  & \\ 
\multicolumn{2}{c}{} & $kT$ (keV) & normalization$^{\S}$ 
                     & $kT$ (keV) & normalization$^{\S}$ 
                     & reduced-$\chi^2$ (d.o.f.)\\ 
\multicolumn{2}{c}{}
  & 0.326$\pm$0.013 & 2.31$\pm$0.37
  & 0.094$\pm$0.004 & 2.02$\pm$0.84 & 1.48 (206)\\ \hline 
\multicolumn{7}{l}{${\ast}$ Parameters of the
 CXB are fixed to the values obtained with HEAO-1 (Boldt et
 al. 1987).}\\
\multicolumn{7}{l}{${\dagger}$ Parameters of the GXE are from the best fit
 result for spectra of the 3.2$^{\circ}$ offset region.}\\
\multicolumn{7}{l}{\hspace{0.3cm}Metal abundance of the APEC model is fixed to 1 solar.}\\
\multicolumn{7}{l}{$\ddagger$ normalization 
= photons ${\rm cm^{-2}}$ s$^{-1}$ keV$^{-1}$
 ($\pi 20^2$ arcmin)$^{-2}$ @1keV. This normalization is estimated} \\
\multicolumn{7}{l}{\hspace{0.3cm}by assuming in the uniform-sky ARF 
 calculation (20' radius).}\\
\multicolumn{7}{l}{$\S$ normalization is for the APEC model and
 equivalent to 10$^{-17}$/\{4$\pi$(1+z)$^2${\it D}$^2_A$\}$\int${\it n}$\rm_e$ {\it n}$\rm_H${\it dV} cm$^{-5}$} \\
\multicolumn{7}{l}{\hspace{0.3cm}({\it D}$_A$ : angular size distance to
 the source).} \\
\end{tabular}
\end{center}
\end{table}

\begin{figure}[h]
\begin{tabular}{c}
\begin{minipage}{1\hsize}
\begin{center}
\rotatebox{-90}{\resizebox{5.0cm}{!}{\includegraphics{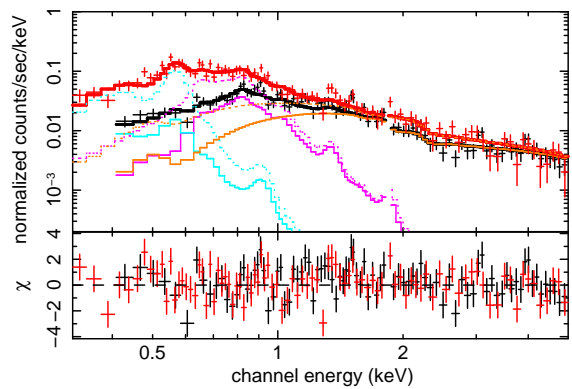}}}
\end{center}
\caption{Best-fit result of spectral analysis of the Galactic emission
 for the 3.2$^{\circ}$ offset observation. Black and red
 crosses are observed spectra of FI and BI, respectively. 
 Black and red thick lines are the best-fit models of FI
 and BI, respectively. As for individual model components, solid lines
 are for FI and dot lines are for BI. Orange is the CXB model, purple and 
 sky-blue are the Galactic emission model for MWH and LHB, respectively. 
 Residuals are shown in the bottom panels.}
\end{minipage}
\end{tabular}
\end{figure}

\subsection{Spectral Fitting}

We fitted the spectra of each region of NGC 4636, based on the
above estimation of the NXB, the CXB, and the GXE. The CXB and the
GXE are included as models in the spectral fit with their spectral
parameters fixed, and the NXB was subtracted.     
We applied a two-component model for the emission from NGC 4636 : 
a single-temperature vAPEC model for the ISM, and a BREMSS model 
with a temperature fixed to 7 keV (Matsushita et al. 1994) for the hard 
component due to low-mass X-ray binaries (LMXB).
For the all regions, the fitting is performed in the energy
range of 0.3--5 keV 
because the spectrum below 0.3 keV is difficult to analyze due to
large systematic errors, and the NXB dominates above 5 keV.
The energy range around the Si K-edge (1.82--1.841 keV) was ignored
due to response problems. 
We divided the metals into several elemental groups 
(He = C = N, O, Ne, Mg = Al, Si, S = Ar = Ca, Fe = Ni), and
fixed the abundance of the He=C=N group to solar. Abundances of
other groups were treated as free parameters. 

\begin{figure}[h]
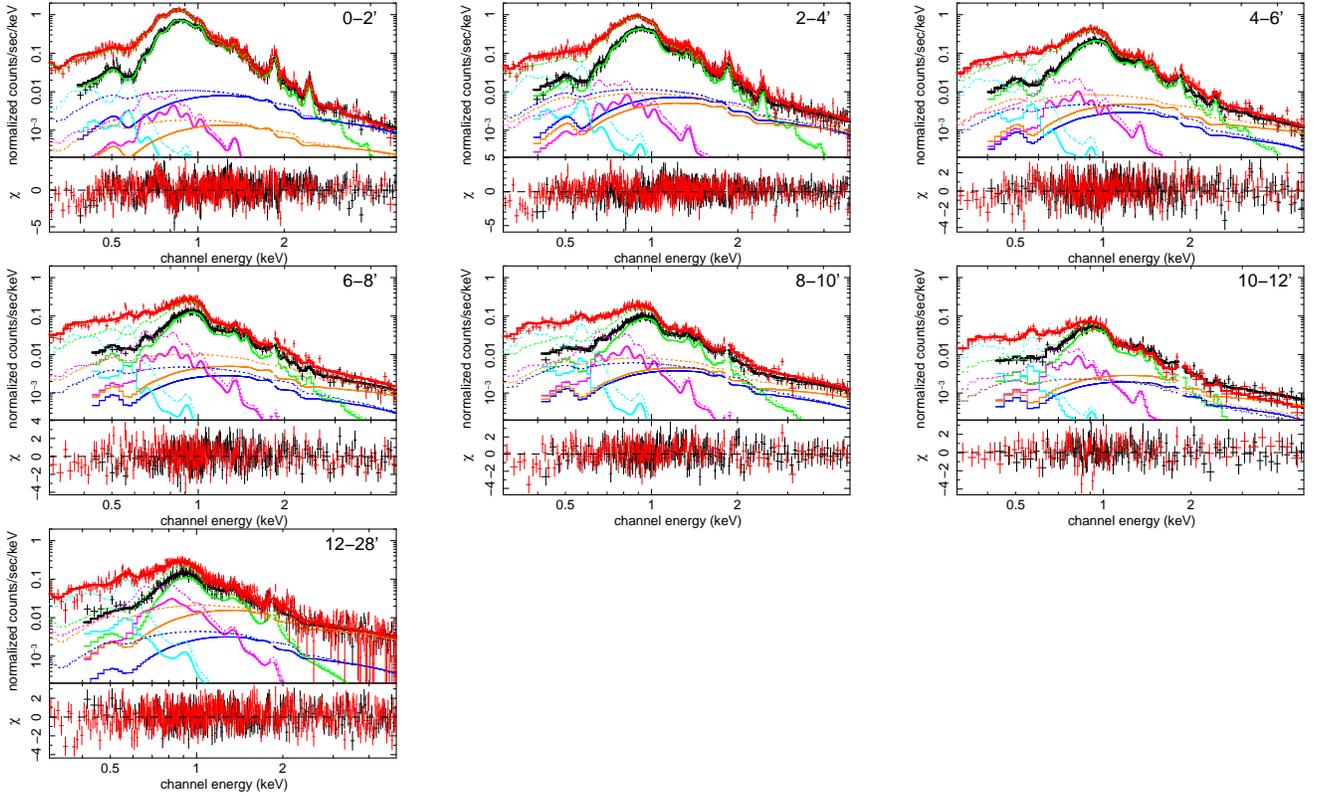

\begin{center}
\begin{tabular}{ccc}
\begin{minipage}{0.33\hsize}
\rotatebox{-90}{\resizebox{3.5cm}{!}{\includegraphics{ngc4636_0_2arcmin_model_ver2.ps}}}
\end{minipage} &
\begin{minipage}{0.33\hsize}
\rotatebox{-90}{\resizebox{3.5cm}{!}{\includegraphics{ngc4636_2_4arcmin_model_ver2.ps}}}
\end{minipage} &
\begin{minipage}{0.33\hsize}
\rotatebox{-90}{\resizebox{3.5cm}{!}{\includegraphics{ngc4636_4_6arcmin_model_ver2.ps}}}
\end{minipage} \\
\begin{minipage}{0.33\hsize}
\rotatebox{-90}{\resizebox{3.5cm}{!}{\includegraphics{ngc4636_6_8arcmin_model_ver2.ps}}}
\end{minipage} &
\begin{minipage}{0.33\hsize}
\rotatebox{-90}{\resizebox{3.5cm}{!}{\includegraphics{ngc4636_8_10arcmin_model_ver2.ps}}}
\end{minipage} &
\begin{minipage}{0.33\hsize}
\rotatebox{-90}{\resizebox{3.5cm}{!}{\includegraphics{ngc4636_10_12arcmin_model_ver2.ps}}}
\end{minipage} \\
\begin{minipage}{0.33\hsize}
\rotatebox{-90}{\resizebox{3.5cm}{!}{\includegraphics{ngc4636_north_0_8arcmin_model_ver2.ps}}}
\end{minipage} 
\end{tabular}
\end{center}
\caption{Spectral fitting of each annular region of NGC 4636.
 Black and red crosses are observed spectra of FI and BI,
 respectively. Black and red thick lines are the best-fit models to the FI
 and BI, respectively. For individual model components, solid lines are 
 for FI and dot lines are for BI. Green and blue are the thermal component and 
 LMXB component from NGC 4636, respectively. Orange is the CXB model, 
 purple and sky-blue are the Galactic emission models for the MWH and LHB, 
 respectively. Residuals are shown in the bottom panels. 
 Shown are spectra of 0--2', 2--4', 4--6' regions in the upper panels, 
6--8', 8--10', 10--12' in the middle panels, 
and 12--28' in the bottom.} 
\end{figure}

\begin{table}[h]
\caption{Results of spectral fitting of each annulus region of NGC 4636}
\begin{center}
\begin{small}
\begin{tabular}{cccccccc} \hline\hline
region   & model  & $kT$ (keV)        & O (solar)     & Ne (solar) & Mg (solar) 
& Si (solar)     & Fe (solar) \\ \hline 
0--2'     & 1T$^a$ & 0.642$\pm$0.002 & 0.75$\pm$0.06 & 1.54$\pm$0.12 & 1.50$\pm$0.11
& 1.51$\pm$0.12 & 1.14$\pm$0.07 \\
0--2'     & 2T$^b$ & 1.56$\pm$0.33,\ 0.641$\pm$0.002 & 0.79$\pm$0.07 & 1.47$\pm$0.15
& 1.52$\pm$0.11 & 1.54$\pm$0.12 & 1.20$\pm$0.07 \\       
2--4'     & 1T   & 0.733$\pm$0.004    & 1.01$\pm$0.13  & 2.14$\pm$0.28 & 2.33$\pm$0.24 
& 1.81$\pm$0.19 & 1.41$\pm$0.12 \\
4--6'     & 1T   & 0.792$\pm$0.005    & 0.41$\pm$0.06  & 0.62$\pm$0.11 & 0.86$\pm$0.08
& 0.57$\pm$0.06 & 0.49$\pm$0.03 \\
6--8'     & 1T   & 0.811$\pm$0.006    & 0.41$\pm$0.08  & 0.45$\pm$0.13 & 0.74$\pm$0.10
& 0.50$\pm$0.07 & 0.43$\pm$0.03 \\
8--10'    & 1T   & 0.796$\pm$0.008    & 0.34$\pm$0.10  & 0.60$\pm$0.17 & 0.86$\pm$0.14
& 0.53$\pm$0.11 & 0.43$\pm$0.05 \\ 
10--12'   & 1T   & 0.774$\pm$0.013    & 0.58$\pm$0.28 & 0.67$\pm$ 0.42 & 1.20$\pm$0.40
& 0.84$\pm$0.39 & 0.68$\pm$0.17 \\
12--28'   & 1T   & 0.746$\pm$0.016    & $<$1.20       & $<$1.60       & 1.13$\pm$0.41  
& 0.87$\pm$0.31 & 0.63$\pm$0.17 \\ \hline
region & \multicolumn{2}{c}{$^c$vAPEC normalization} 
       & \multicolumn{3}{c}{$^d$BREMSS normalization}
       & \multicolumn{2}{c}{reduced-$\chi^2$ (d.o.f.)} \\ \hline
0--2'   & \multicolumn{2}{c}{2.12$\pm$0.21} 
       & \multicolumn{3}{c}{9.58$\pm$1.01} 
       & \multicolumn{2}{c}{1.40 (829)}\\
0--2'   & \multicolumn{2}{c}{0.10$\pm$0.06, 2.01$\pm$0.20} 
       & \multicolumn{3}{c}{7.73$\pm$2.13} 
       & \multicolumn{2}{c}{1.38 (827)}\\
2--4'   & \multicolumn{2}{c}{1.08$\pm$0.17} 
       & \multicolumn{3}{c}{9.34$\pm$1.30} 
       & \multicolumn{2}{c}{1.29 (844)}\\
4--6'   & \multicolumn{2}{c}{1.51$\pm$0.14} 
       & \multicolumn{3}{c}{3.98$\pm$1.17} 
       & \multicolumn{2}{c}{1.11 (518)}\\
6--8'   & \multicolumn{2}{c}{0.97$\pm$0.13} 
       & \multicolumn{3}{c}{4.04$\pm$1.12} 
       & \multicolumn{2}{c}{1.08 (446)}\\
8--10'  & \multicolumn{2}{c}{0.72$\pm$0.13} 
       & \multicolumn{3}{c}{5.34$\pm$1.19} 
       & \multicolumn{2}{c}{1.01 (371)}\\
10--12' & \multicolumn{2}{c}{0.27$\pm$0.13} 
       & \multicolumn{3}{c}{3.18$\pm$1.21} 
       & \multicolumn{2}{c}{0.96 (210)}\\
12--28' & \multicolumn{2}{c}{0.65$\pm$0.33} 
       & \multicolumn{3}{c}{3.60$\pm$2.36}             
       & \multicolumn{2}{c}{1.01 (586)}\\ \hline
\multicolumn{8}{l}{$^a$\ single-temperature vAPEC model}\\
\multicolumn{8}{l}{$^b$\ two-temperature vAPEC model}\\
\multicolumn{8}{l}{$^c$\ [10$^{-17}$/\{4$\pi$(1+z)$^2${\it
 D}$^2_A$\}]$\int${\it n}$\rm_e${\it n}$\rm_H${\it dV} cm$^{-5}$}\\
\multicolumn{8}{l}{\hspace{0.3cm}({\it n}$\rm_e$ : electron density, {\it n}$\rm_H$ :
 hydrogen density, {\it D}$_A$ : angular size distance to the source)}\\
\multicolumn{8}{l}{$^d$\{3.02$\times$10$^{-20}$/4$\pi${\it
 D}$^2$\}$\int${\it n}$\rm_e${\it n}$\rm_I${\it dV} cm$^{-5}$ ({\it
 n}$\rm_I$ : ion density, {\it D} : distance to the source)}\\
\end{tabular}
\end{small}
\end{center}
\end{table}

Table 3 shows the best-fit results for each region. The fit is  
acceptable for all the regions.
The O abundance is measured up to 10--12' and Mg, Si and Fe
to 12--28' for the first time.
However, the Ne abundance is not reliable because Ne line measurements are influenced
by nearby Fe-L lines, which cannot be resolved with the resolution of the XIS.
It can be seen from Figure 3 
that the Galactic emission is well reproduced and 
significantly contributes to the spectra 
in the outer region, and thus it must be accurately estimated to derive the O
abundance of NGC 4636.
As a result, even for O, the abundances in the outer regions
are determined up to 10' with $\sim$30\% accuracy.    
Even if we allow the temperature
of the BREMSS (7keV) component to vary, the metal abundances only
change within errors so we believe that they are reliable. 
The luminosity of the hard 
X-ray component of the BREMSS model is 1.5$\times$10$^{40}$ erg s$^{-1}$
over 2--10 keV within 6' of NGC 4636. 
This is consistent with the ASCA results 
(Matsushita et al. 1994), indicating that 
the hard X-ray component is due to LMXB.
We also confirmed that the abundance results are almost the same
when we fit the hard component with a power-law of index 1.5, which was
used for the LMXB estimation in Loewenstein et al. (2001).
In the central region, the reduced-$\chi^2$ is a little larger than that in
the outer region.
This might be due to temperature structure in the center region
(Ohto et al. 2003) so we tried fitting with a
two-temperature model only in the 0--2' region. 
We assumed that the metal abundances
are the same for both temperature components. Comparing 
the result with that of the single-temperature model,
the metal abundances do not greatly change 
and the temperature of the hot component in the two-temperature model is almost the
same as that in the single-temperature.
The F-test probability is 0.097\% and thus the improvement is marginal.
Therefore, hereafter we  
focus on the results with the single-temperature model.
A fit residual around 0.7 keV in the 0--2' region is discussed
and analyzed in \S3.4.

\subsection{Temperature and Metal Abundances}

Distributions of temperature and metal abundances are plotted 
against radius (kpc) 
in Figures 4 and 5, respectively. Systematic errors due to the GXE, CXB,
and NXB are also shown.
The systematic errors due to the GXE 
are estimated by varying the normalizations of
the APEC model for the MWH and LHB within their error ranges, which are
comparable to the X-ray count fluctuation (15\%) 
of the ROSAT All-sky X-ray background
survey (band 4)\footnote{SkyView: {\tt http://skyview.gsfc.nasa.gov/}}
for the NGC 4636 region.
The systematic errors due to the CXB and NXB are at most 5\%
(Tawara et al. 2008), respectively, and they are dominant in the hard
X-ray band.
Therefore, the systematic errors on the temperature and metal abundances
are mostly caused by the GXE uncertainty.
A 15\% uncertainty in the estimate of the XIS optical blocking filter
thickness introduces systematic errors comparable to those from the GXE.
The temperature in the central region is 0.64 keV, rising to 0.7--0.8 keV.
This profile is consistent with that from ASCA (Matsushita et al. 1997), XMM-Newton
(Xu et al. 2002) and Chandra (Ohto et al. 2003)  
and the temperature distribution is barely affected by any systematic
errors in the GXE.
For the inner 50 kpc most metal abundances are tightly restricted
thanks to the high energy resolution of the Suzaku XIS while in the
outer regions the abundances are less well constrained due to the GXE
systematic errors.
However, the Ne abundance is not reliable because Ne lines are confused
with Fe-L lines.
Furthermore, only upper limits could be obtained on the O and Ne abundances in 
the 
60--140 kpc region, since their uncertainties are very large 
due to contamination by the Galactic emission. 
It can been seen that
all the metal abundances are 0.7--2.5 solar in the central region, and
decrease down to 0.3--1.0 solar towards the outer region.
In the central region, the O and Fe abundances are about 1.2 times larger
than that obtained with the RGS (Xu et al. 2002).
Figure 6 shows confidence contours of O and Mg abundances against the Fe
abundance.
The O and Mg to Fe abundance ratio are particularly important to constrain
the contribution of different types of supernovae.
Although the individual errors for each metal
abundance are large, the abundance ratios are constrained well, 
especially for the Mg to Fe abundance ratio in the
outermost region.

\begin{figure}[h]
\begin{tabular}{c}
\begin{minipage}{1\hsize}
\begin{center}
\rotatebox{-90}{\resizebox{5cm}{!}{\includegraphics{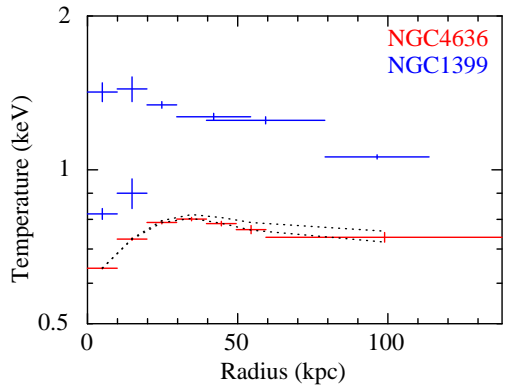}}}
\end{center}
\caption{Temperature distribution of NGC 4636 (red) plotted against the
radius in units of kpc. Dotted lines indicate the systematic error range 
due to the GXE uncertainty.
For comparison, Suzaku results of NGC 1399
 (blue) (Matsushita et al. 2007a) are also shown. }
\end{minipage}
\end{tabular}
\end{figure}

\begin{figure}[h]
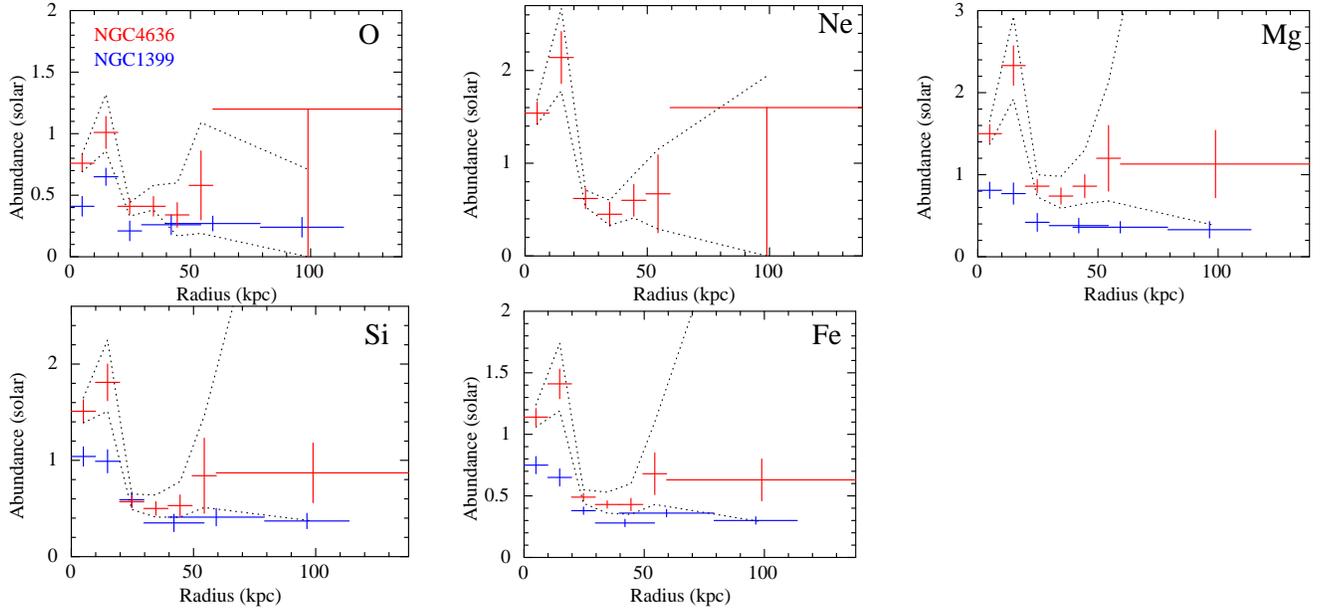

\begin{center}
\begin{tabular}{ccc}
\begin{minipage}{0.33\hsize}
\rotatebox{-90}{\resizebox{4cm}{!}{\includegraphics{NGC1399_NGC4636_O.ps}}}
\end{minipage} &
\begin{minipage}{0.33\hsize}
\rotatebox{-90}{\resizebox{4cm}{!}{\includegraphics{NGC4636_Ne.ps}}}
\end{minipage} &
\begin{minipage}{0.33\hsize}
\rotatebox{-90}{\resizebox{4cm}{!}{\includegraphics{NGC1399_NGC4636_Mg.ps}}}
\end{minipage} \\
\begin{minipage}{0.33\hsize}
\rotatebox{-90}{\resizebox{4cm}{!}{\includegraphics{NGC1399_NGC4636_Si.ps}}}
\end{minipage} &
\begin{minipage}{0.33\hsize}
\rotatebox{-90}{\resizebox{4cm}{!}{\includegraphics{NGC1399_NGC4636_Fe.ps}}}
\end{minipage} 
\end{tabular}
\end{center}
\caption{Metal abundance distributions of NGC 4636 (red) plotted against
the radius in units of kpc. Dotted lines indicate the systematic error range 
due to the GXE uncertainty.
For comparison,
Suzaku results of NGC 1399
 (blue) (Matsushita et al. 2007a) are also shown. The O and Ne abundances in
 60--140 kpc
represent an upper limit.} 
\end{figure}

\begin{figure}[h]
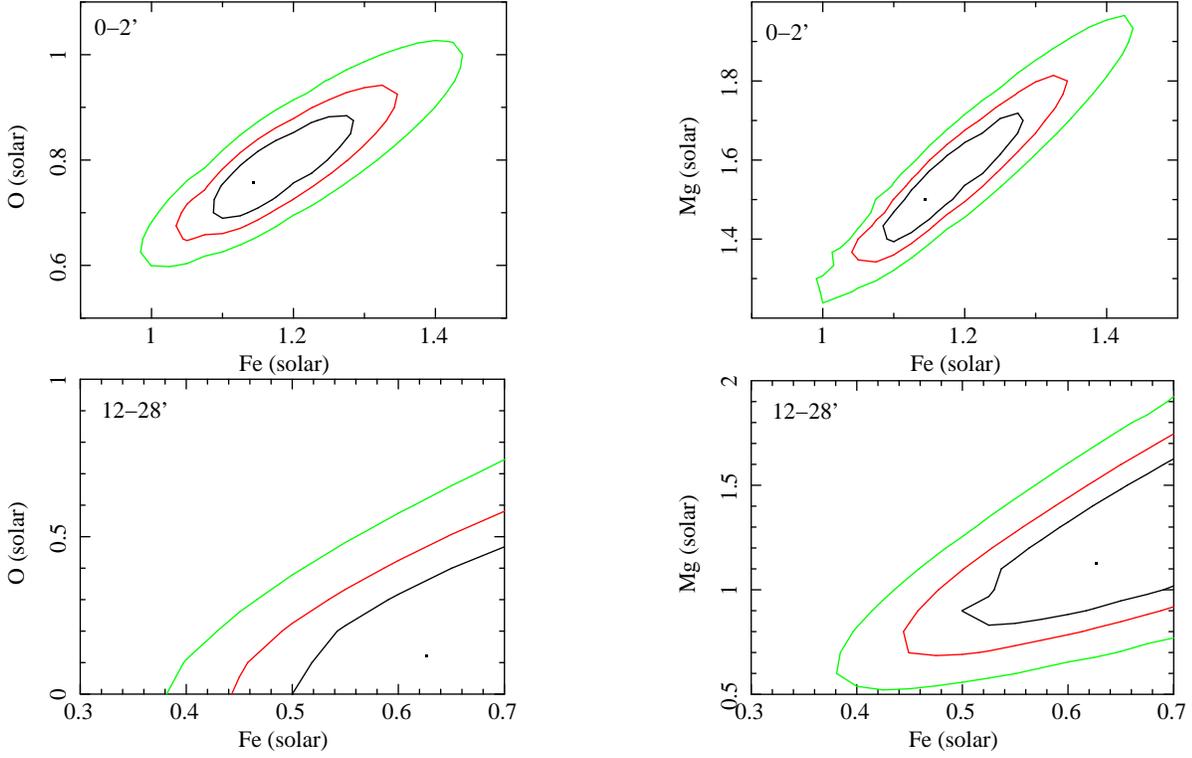

\begin{center}
\begin{tabular}{cc}
\begin{minipage}{0.5\hsize}
\rotatebox{-90}{\resizebox{5cm}{!}{\includegraphics{Fe_O_cont_0_2arcmin.eps}}}
\end{minipage} &
\begin{minipage}{0.5\hsize}
\rotatebox{-90}{\resizebox{5cm}{!}{\includegraphics{Fe_Mg_cont_0_2arcmin.eps}}}
\end{minipage} \\
\begin{minipage}{0.5\hsize}
\rotatebox{-90}{\resizebox{5cm}{!}{\includegraphics{Fe_O_cont_0_8arcmin.eps}}}
\end{minipage} &
\begin{minipage}{0.5\hsize}
\rotatebox{-90}{\resizebox{5cm}{!}{\includegraphics{Fe_Mg_cont_0_8arcmin.eps}}}
\end{minipage} 
\end{tabular}
\end{center}
\caption{Confidence contours of Fe abundance vs O abundance (left), and Fe
 abundance vs Mg abundance (right) for the inner (top) and outer (bottom)
 region. The square mark denotes the best-fit location,
 and the contours represent 68\%, 90\%, 99\% confidence level, respectively.}
\end{figure}

\subsection{Resonance Scattering in the Central Region}

In Figure 7, we show the spectra in the 0--2'and 2--4' regions. 
In the spectrum of the 0--2' region,
the fit residual has a negative line feature around 0.85 keV. 
This feature disappears in the
2--4' region and other outer regions (Figure 3). 
This is possibly due to resonance scattering 
of the Fe ${\rm_{XV\hspace{-.1em}I\hspace{-.1em}I}}$ line, which was 
reported with the RGS (Xu et al. 2002; Werner et al. 2009).
When we introduce a negative gaussian line in the spectral model, the
fit significantly improves with a F-test probability of $<0.01$\%.
In that case, the Fe abundance increases by 15\%.
We compared the 
XIS result in the central region with that using the RGS and 
looked into whether the features are consistent. 
We modeled the XIS spectrum with two 
BREMSS for the continuum thermal
emission from the plasma and the LMXBs, and
add each emission line with a gaussian. 
The temperatures of these BREMSS
models are fixed to 0.642 keV and 7.0 keV for the thermal and LMXB emission,
respectively. 
The former is the best-fit vAPEC temperature in table 3. 
The lines considered here are listed in Table 4.
Since the XIS cannot resolve individual lines,
we summed the flux of neighboring lines within 30 eV in the RGS paper
(Xu et al. 2002),
and assumed the strength-weighted line energy. 
Furthermore, we refer to the ATOMDB line table
\footnote{\tt http://cxc.harvard.edu/atomdb/WebGUIDE/index.html} 
for the energy and intensity of the 0.961 and 0.963 keV lines, 
which were not analyzed in the RGS spectrum, but are significant in the XIS
spectra. 
The Ne lines around this energy range are not shown because their fluxes
are not reliable due to the influence of Fe-L lines. 
The line width of all lines is fixed to 0 keV.

The line intensity ratio 
against the summed flux of the strongest line (0.737 and 0.728 keV ) of 
Fe ${\rm_{XV\hspace{-.1em}I\hspace{-.1em}I}}$
is summarized in Table 4 and Figure 8. 
In Figure 8,
the predicted line intensity ratios are also shown. These 
were obtained by simulating spectra with the emission models and 
parameters from Table 3 
and fitting them with the same model used for the data.
The result indicates that the flux intensity of the 
Fe ${\rm_{XV\hspace{-.1em}I\hspace{-.1em}I}}$ line (0.825 keV),
which was suggested to be affected by resonance scattering (Xu et al. 2002),
is about 0.75 times lower than that predicted by the model.
This might indicate that the resonance scattering occurs for the 
Fe ${\rm_{XV\hspace{-.1em}I\hspace{-.1em}I}}$ ion.
The Suzaku intensity ratio is about 1.4 times higher than that from the RGS.
This might be because the RGS photons are accumulated from a smaller
central region than the XIS where the resonance scattering is more
effective. 
Doron and Behar (2002) indicated the problem of modeling the Fe+16/Fe+17
ion lines, but their recalculation showed that the line intensity ratio
of the 0.825 keV line to the sum of the 0.737 and 0.728 keV lines
changes by 30--40\%, mostly independent of the temperature in the range
of 0.6--0.9 keV.
Their paper does not refer to the APEC model, and thus this factor of
change is not necessarily accepted for our cases.
Our results show that the discrepancy of the line intensity ratio
between the data and the model is seen only at the central region.
Therefore, we conclude that the resonance scattering is robust.

\begin{figure}[h]
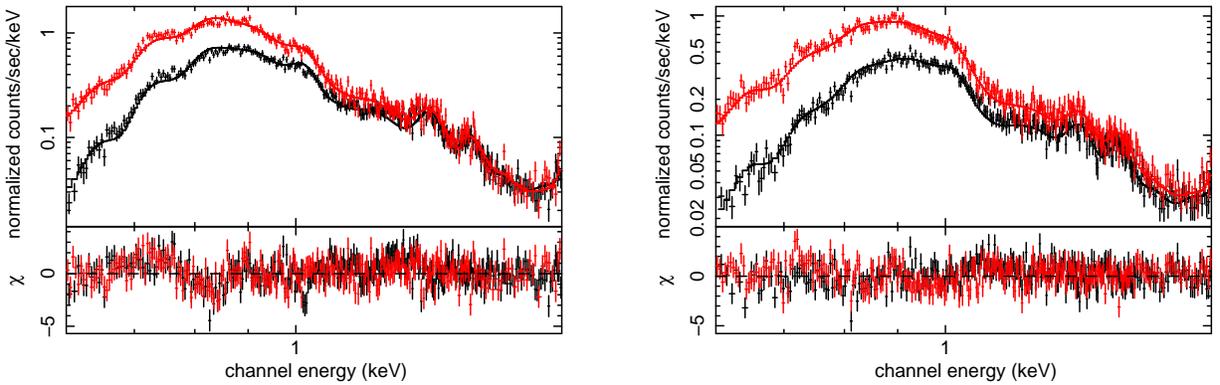

\begin{center}
\begin{tabular}{cc}
\begin{minipage}{0.5\hsize}
\begin{center}
\rotatebox{-90}{\resizebox{5.0cm}{!}{\includegraphics{ngc4636_0_2arcmin_resonance_scattering.ps}}}
\end{center}
\end{minipage}
\begin{minipage}{0.5\hsize}
\begin{center}
\rotatebox{-90}{\resizebox{5.0cm}{!}{\includegraphics{ngc4636_2_4arcmin_resonance_scattering.ps}}}
\end{center}
\end{minipage}
\end{tabular}
\end{center}
\caption{Spectrum of 0--2'(left) and 2--4'(right) region. 
 Black and red
 crosses are observational spectra of FI and BI, respectively. 
 Black and red lines are the best-fit models of FI
 and BI, respectively.
 Residuals 
 are shown in the bottom panels.}
\end{figure}

\begin{table}
\caption{Summary of line intensities in 0.4--1.5 keV for the RGS (Xu et
 al. 2002) and the XIS.
Line intensity ratio is against the sum flux of 0.737 keV and 0.728 keV Fe lines. }
\begin{center}
\begin{small}
\begin{tabular}{cccccccc} \hline\hline 
Ion     & \multicolumn{4}{c}{RGS} 
                & \multicolumn{3}{c}{XIS} \\ 
        & $E$ (keV) & *flux & *sum of flux & intensity ratio  
            & $E$ (keV) & *flux & intensity ratio \\ \hline
N ${\rm_{V\hspace{-.1em}I}}$ & 0.418-0.428 & 0.43 & 0.43 & 0.05 & 0.428 & 0 & 0 \\ \hline
N ${\rm_{V\hspace{-.1em}I\hspace{-.1em}I}}$ & 0.497       & 0.20 & 0.20 & 0.03 & 0.497 & 0 & 0 \\ \hline  
O ${\rm_{V\hspace{-.1em}I\hspace{-.1em}I}}$ & 0.557       & 0.21 &      &      &       &   & \\ 
O ${\rm_{V\hspace{-.1em}I\hspace{-.1em}I}}$ & 0.565       & 0.09 &      &      &       &   & \\ 
O ${\rm_{V\hspace{-.1em}I\hspace{-.1em}I}}$ & 0.570       & 0.13 & 0.43 & 0.05 & 0.563 & 0 & 0 \\ \hline
O ${\rm_{V\hspace{-.1em}I\hspace{-.1em}I\hspace{-.1em}I}}$ & 0.651       & 1.96 & 1.96 & 0.245 & 0.651 
       & 1.03$\pm$0.18 & 0.21$\pm$0.04  \\ \hline    
Fe ${\rm_{XV\hspace{-.1em}I\hspace{-.1em}I}}$ & 0.728     & 5.66 &      &      &       &   & \\ 
Fe ${\rm_{XV\hspace{-.1em}I\hspace{-.1em}I}}$ & 0.737     & 2.33 & 7.99 & 1    & 0.731  
       & 4.85$\pm$0.26 & 1.00$\pm$0.08  \\ \hline 
Fe ${\rm_{XV\hspace{-.1em}I\hspace{-.1em}I\hspace{-.1em}I}}$  
Fe ${\rm_{XI\hspace{-.1em}X}}$      
O ${\rm_{V\hspace{-.1em}I\hspace{-.1em}I\hspace{-.1em}I}}$   & 0.769     & 1.99 & 1.99 & 0.249 & 0.769 
       & 2.94$\pm$0.28 & 0.61$\pm$0.07 \\ \hline 
Fe ${\rm_{XV\hspace{-.1em}I\hspace{-.1em}I}}$ & 0.809     & 1.66 &      &       &      &   & \\ 
Fe ${\rm_{XV\hspace{-.1em}I\hspace{-.1em}I}}$ & 0.825     & 3.56 & 5.22 & 0.653 & 0.820 
       & 4.46$\pm$0.26 & 0.92$\pm$0.07 \\ \hline
Fe ${\rm_{XV\hspace{-.1em}I\hspace{-.1em}I\hspace{-.1em}I}}$ & 0.853     & 2.25 &      &       &      &   & \\ 
Fe ${\rm_{XV\hspace{-.1em}I\hspace{-.1em}I\hspace{-.1em}I}}$ & 0.859     & 1.85 &      &       &      &   & \\ 
Fe ${\rm_{XV\hspace{-.1em}I\hspace{-.1em}I\hspace{-.1em}I}}$
Fe ${\rm_{XX}}$ & 0.865     & 1.45 &      &       &      &   & \\ 
Fe ${\rm_{XV\hspace{-.1em}I\hspace{-.1em}I\hspace{-.1em}I}}$ & 0.871     & 1.90 & 7.45 & 0.932 & 0.861 
       & 4.98$\pm$0.23 & 1.03$\pm$0.07 \\ \hline
Fe ${\rm_{XX}}$ & 0.961     & -    &      &       &      &   & \\ 
Fe ${\rm_{XX}}$ & 0.963     & -    & -    &  -    & 0.962 
       & 1.73$\pm$0.15 & 0.36$\pm$0.04 \\ \hline
Mg ${\rm_{XI}}$ & 1.35      & 0.59 & 0.59 & 0.070 & 1.35 
       & 0.61$\pm$0.04 & 0.13$\pm$0.01 \\ \hline 
\multicolumn{8}{l}{*flux : 10$^{-4}$photons cm$^{-2}$ s$^{-1}$}\\
\end{tabular}
\end{small}
\end{center}
\end{table}

\begin{figure}[h]
\begin{tabular}{c}
\begin{minipage}{1\hsize}
\begin{center}
\rotatebox{-90}{\resizebox{6.0cm}{!}{\includegraphics{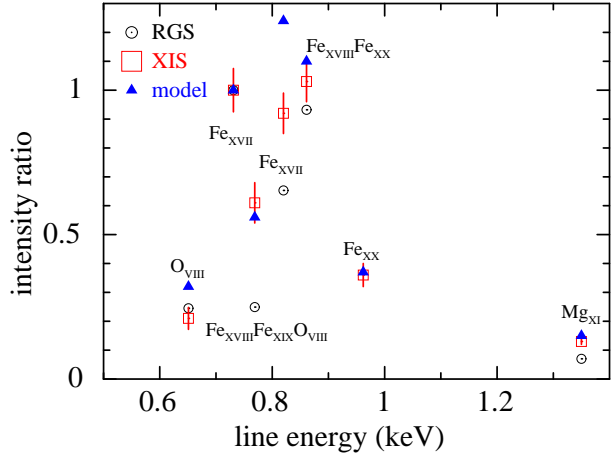}}}
\end{center}
\caption{Relative line intensity against the sum flux of 0.733
 keV and 0.728 keV lines of Fe
 ${\rm_{XV\hspace{-.1em}I\hspace{-.1em}I}}$.
Squares and circles are the data of the XIS and RGS (Xu et al. 2003).
The horizontal axis represents line energy, and the vertical axis
 relative line intensity. Model prediction (filled triangle) is 
obtained by simulating spectra with the emission models and
parameters on Table 3 and fitting them with BREMSS and ZGAUSS models.}
\vspace{-1.0em}
\end{minipage}
\end{tabular}
\end{figure}    

\section{Discussion}

We measured the temperature and metal abundance
distributions of NGC 4636 out to 28' and confirmed the resonance scattering
in the central region. In this section we compare the Suzaku results 
with those for other X-ray bright elliptical galaxies (NGC 1399,
NGC 507, and NGC 5044), 
and discuss the metal enrichment process in elliptical galaxies. 

\subsection{Metallicity Distribution}

In Figures 4 and 5, observational results by the Suzaku XIS  
for NGC 1399 (Matsushita et al. 2007a) are also shown. 
These are the results with a two-temperature model
within 4' and one-temperature model beyond 4'.
The temperature of NGC 1399 is about twice as high as that of NGC 4636, 
indicating that the gravitational potential of NGC 1399 is deeper.
In Figure 5, the Ne
abundance of NGC 1399 was not measured because its line was polluted by
Fe-L lines. The radial distributions of the O, Mg,
Si and Fe abundances are similar between the two galaxies, but
metal abundances are about twice as high as
those of NGC 1399 in all radial regions. 
In Figure 9, radial profiles of the abundance ratios of O, Ne, Mg, and Si
against Fe are shown, with error bars estimated from the allowed
slope of the confidence contours in Figure 6.
For comparison, the abundance ratios for a model of SN II (for all
elements)  
and SN Ia (for Si) are also shown.
The abundance ratios for SN II are 
calculated by integrating the metal mass ejected by
SNe II (Iwamoto et al. 2006) with a progenitor mass from 10M$_{\odot}$ 
to 50M$_{\odot}$, weighting by
the Salpeter initial mass function, and calculated, for example, as 
O: 1.80M$_{\odot}$ and Fe: 0.0907M$_{\odot}$. 
The W7 model
(O: 0.143M$_{\odot}$ and Fe: 0.749M$_{\odot}$ ; Iwamoto et al. 1999)
is used for SN Ia. 
The result indicates that all metal
abundances can be explained by a mixture of 
SN II and SN Ia products. 
This indicates that, in NGC 4636
and NGC 1399, each metal diffused to the outside of the galaxy after mixing 
products from both SN II and SN Ia. 
For NGC 4636, all the abundance ratios are
constant in the radial direction. The abundance
ratios against Fe in the central region (0--10 kpc) are 
        0.66$\pm^{0.07}_{0.06}$ solar, 1.35$\pm^{0.12}_{0.10}$ solar, 
        1.32$\pm^{0.06}_{0.07}$ solar, 1.32$\pm^{0.11}_{0.10}$ solar for 
O, Ne, Mg, Si, respectively. 
The O/Fe ratio is similar to that of  
NGC 1404 (Matsushita et al. 2007a) and NGC 720 (Tawara et al. 2006) measured by
the XIS, and also with the RGS results for
NGC 4636 (Xu et al. 2002) and NGC 5044 (Tamura et al. 2003). This indicates 
that a common metal production process has taken
place in elliptical galaxies. However, the Mg/Fe ratio of NGC 4636 is
higher than that of NGC 1399 in the outer region.
This difference might be caused by different populations of SN II
progenitor stars between the two galaxies, but problems with the plasma emission
model cannot be ruled out (Matsushita et al. 2003).

It is noticeable that the radial profile of the O abundance of NGC 4636
exhibits a steeper gradient than other elliptical galaxies, NGC 1399 
, NGC 5044, and NGC 507.
This indicates that NGC 4636 keeps metals more tightly concentrated.
NGC 4636 is rather isolated and has never been identified as a galaxy group; 
the other galaxies are classified as the center galaxy of their galaxy group,
which contains many dwarf galaxies (Ferguson and Sandage 1990; Tifft
et al. 1975).
Therefore, the differences in the abundance gradient could be explained
by NGC 4636 experiencing fewer galaxy interactions leading to reduced
diffusion of metals to the outer region. We consider this further in
the next section.

\begin{figure}[h]
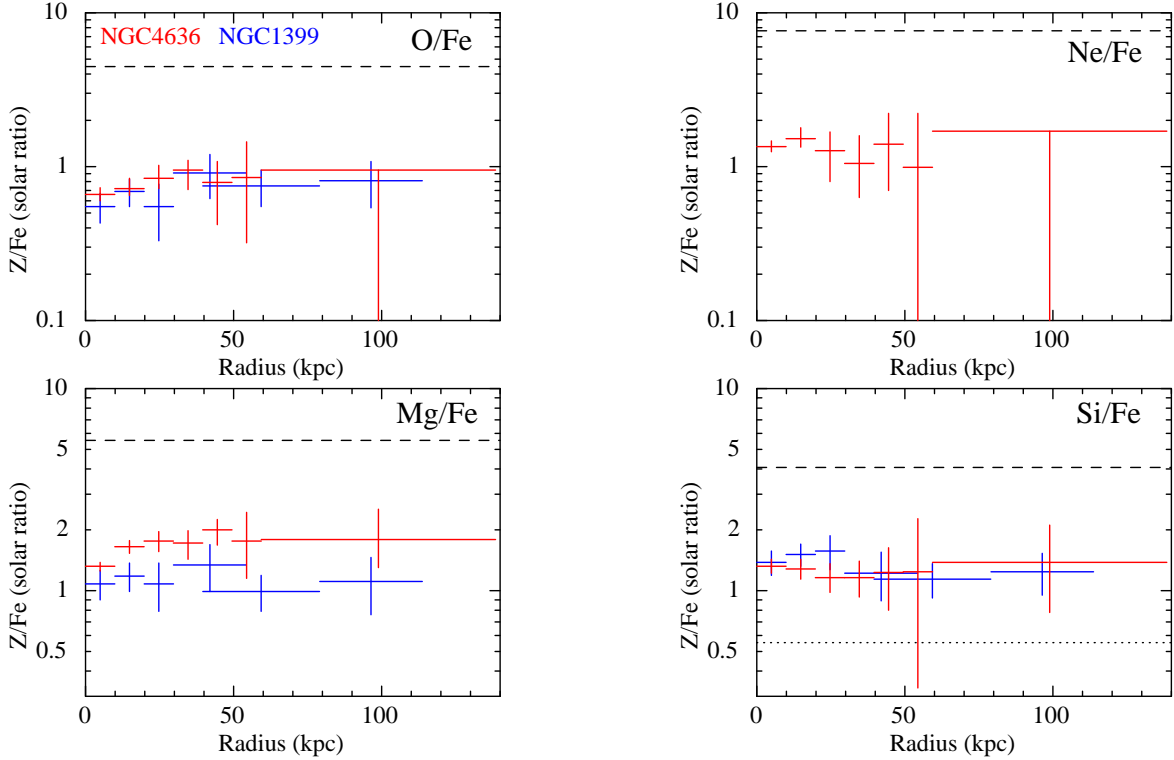

\begin{center}
\begin{tabular}{cc}
\begin{minipage}{0.5\hsize}
\rotatebox{-90}{\resizebox{5cm}{!}{\includegraphics{NGC4636_NGC1399_abund_ratio_Fe_O.ps}}}
\end{minipage} &
\begin{minipage}{0.5\hsize}
\rotatebox{-90}{\resizebox{5cm}{!}{\includegraphics{NGC4636_abund_ratio_Fe_Ne.ps}}}
\end{minipage} \\
\begin{minipage}{0.5\hsize}
\rotatebox{-90}{\resizebox{5cm}{!}{\includegraphics{NGC4636_NGC1399_abund_ratio_Fe_Mg.ps}}}
\end{minipage} &
\begin{minipage}{0.5\hsize}
\rotatebox{-90}{\resizebox{5cm}{!}{\includegraphics{NGC4636_NGC1399_abund_ratio_Fe_Si.ps}}}
\end{minipage} 
\end{tabular}
\end{center}
\caption{Radial profiles of metals to Fe abundance ratios 
for NGC 4636 (red) and NGC 1399 (blue; Matsushita et al. 2007a).
The dashed line represents the ratios for 
the SN II model (Nomoto et al. 2006), and the dotted line the SN Ia model 
(Iwamoto et al. 1999). The O, Ne and Mg abundance ratios for
SN Ia are not shown because these values are quite small ($\simeq$
 0.01). O and Ne abundances in
 60--140 kpc
are upper limits.} 
\end{figure}

\subsection{Metal-mass-to-light-ratio}

The gas mass distributions of NGC 4636 and NGC 1399 are shown in Figure 10, 
which also shows the Chandra results within 0.022 $r_{180}$ 
($\simeq$ 15.6 kpc) (Fukazawa et al. 2006) and the ASCA 
results beyond this radius (Matsushita et al. 1998).
These results show that the gas mass of NGC 4636 is about half as much as 
that of NGC 1399. This is possibly because 
NGC 1399 sits at the center of the Fornax
cluster, where the potential is deeper than that for NGC 4636 (Figure 4),
so the gas is more tightly bound by NGC 1399. 

The NGC 4636 metal mass distributions and metal-mass-to-light-ratios
(MLR) of O and Fe
are shown in Figures 11(a) and (b), respectively. The metal mass is obtained 
by multiplying the metal abundance value in the outermost region in
Figure 5 with the gas mass distribution (Figure 10).
The stellar light distribution is quoted from Fukazawa et al. (2006).

\begin{figure}[h]
\begin{tabular}{c}
\begin{minipage}{1\hsize}
\begin{center}
\rotatebox{-90}{\resizebox{6.0cm}{!}{\includegraphics{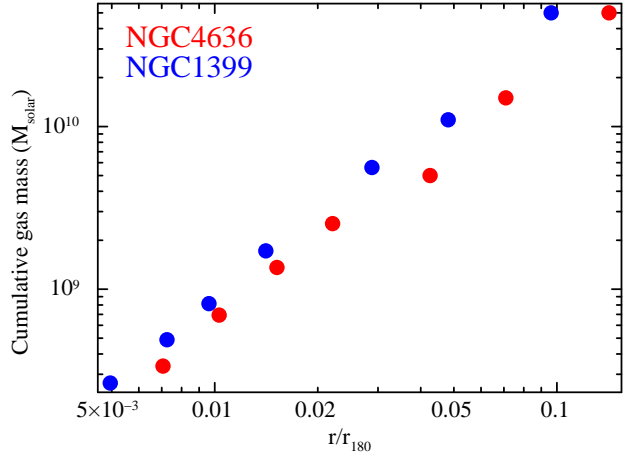}}}
\end{center}
\caption{Gas mass distributions of NGC 4636 (red) and NGC 1399 (blue).
 For NGC 4636, data within and beyond 0.022 $r_{180}$ ($\simeq$ 15.6 kpc)  
 refer to the results of Chandra (Fukazawa et al. 2006) and ASCA 
 (Matsushita et al. 1998), respectively.
 For NGC 1399, data within and beyond 0.014 $r_{180}$ ($\simeq$ 14.7 kpc) 
 refer to the results
of Chandra (Fukazawa et al. 2006) and
ASCA (Ikebe 1995) respectively.
Double-$\beta$ model is used as a gas density distribution.}
\end{minipage}
\end{tabular}
\end{figure}

\begin{figure}[h]
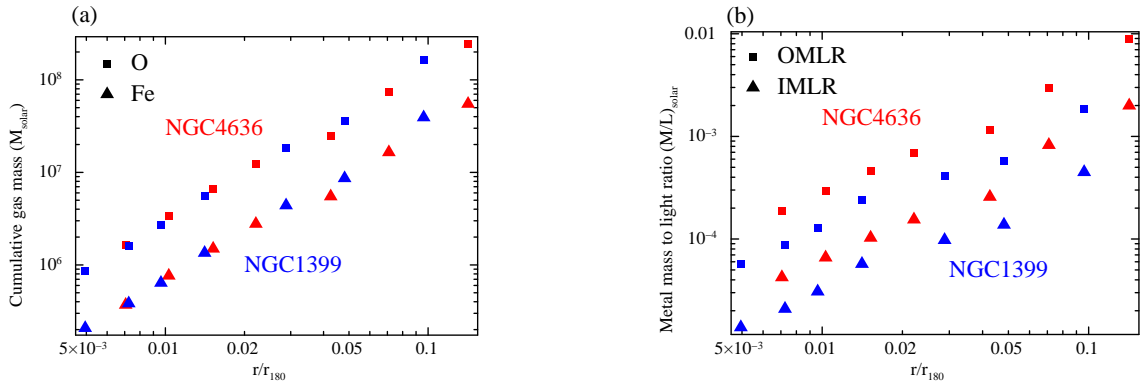

\begin{center}
\begin{tabular}{cc}
\begin{minipage}{0.5\hsize}
\rotatebox{-90}{\resizebox{5.0cm}{!}{\includegraphics{NGC4636_NGC1399_O_Fe_mass.ps}}}
\end{minipage}
\begin{minipage}{0.5\hsize}
\rotatebox{-90}{\resizebox{5.0cm}{!}{\includegraphics{NGC4636_NGC1399_OMLR_IMLR.ps}}}
\end{minipage}
\end{tabular}
\end{center}
\caption{(a) Mass distributions of O and Fe in the hot gas of NGC 4636 and
NGC 1399. 
(b) Metal-mass-to-light-ratios of NGC 4636 and NGC 1399. 
The stellar light distribution is quoted from Fukazawa et al. (2006).
 OMLR and IMLR represents oxygen-mass-to-light ratio and
 iron-mass-to-ligh ratio, respectively.
}
\end{figure}

Figure 11(a) shows that the O and Fe mass distributions are similar between
the two galaxies and the metal mass of NGC 4636 is almost the same  
as that of NGC 1399 at all radii. 
Combining this with the difference of the gas mass, it can be
understood that the abundance and
the gas mass of NGC 4636 are about twice and half those of NGC
1399, respectively. 
As seen in Figure 11(b), 
MLR of NGC 4636 is also 2$\sim$3 times larger than 
that of NGC 1399 in all regions for both metals.
This indicates that the metals in the ISM are distributed more compactly in 
NGC 4636; the metals in NGC 4636 do not diffuse so widely.
This difference might be caused by environmental effects; galaxy
interaction is frequent for NGC 1399 and thus metal-rich gas could
be stirred by galaxy motions.


Finally, we compared the MLR of NGC 4636 with Suzaku results of other 
groups and clusters.
These objects are galaxy groups centered on NGC 1399, NGC 5044, and
NGC 507.
They are X-ray bright elliptical galaxies like NGC 4636, but the galaxy
density is high around them and thus they are identified as centers of galaxy
groups.
We also compare with galaxy clusters such as Abell
262, Abell 1060, AWM 7, and Centaurus. 
Following Sato at al. (2009b),
we plot the correlation between
temperature and MLR at 0.1 $r_{180}$ in figure 12. 
This result shows that
MLR of groups and galaxies with a temperature of $\sim$1 keV, 
including NGC 4636,
is $\sim$5 times lower than that for clusters. 
Among galaxies and groups, 
NGC 4636, NGC 5044 and HCG 62 exhibit a spherically symmetric X-ray emission,
and NGC 4636 has the lowest MLR among them. 
A possible correlation between  
temperature and MLR is suggested only for these spherically symmetric objects.
Such a correlation could be explained by higher temperature systems
possessing a deeper gravitational potential which prevents dispersal
of the metals.
Although galaxy velocity dispersion is larger for higher temperature
clusters, there are no big galaxies around cD galaxies.
Therefore, we think that the deep potential can confine the metals
even if a small galaxy passes around the cD galaxy.
On the other hand, the X-ray asymmetric groups NGC 1399 and NGC 507 show a 
relatively low MLR.
These imply that NGC 1399 and NGC 507 experience galaxy interactions
frequently and  
metal-rich gas could be stirred by galaxy motions.

\begin{figure}[h]
\begin{tabular}{c}
\begin{minipage}{1\hsize}
\begin{center}
\rotatebox{-90}{\resizebox{7.0cm}{!}{\includegraphics{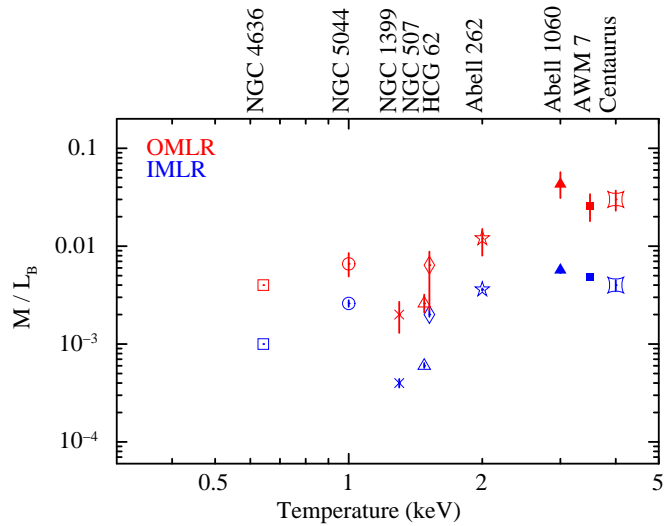}}}
\end{center}
\caption{IMLR and OMLR at 0.1 $r_{180}$ using B band luminosity.
References : NGC 5044 (Komiyama et al. 2008), NGC 1399 (Matsushita et al. 2007a), 
NGC 507 (Sato et al. 2009a), HCG 62 (Tokoi et al. 2008), Abell 262 
(Sato et al. 2009b), Abell 1060 (Sato et al. 2007a), 
AWM 7 (Sato et al. 2008), Centaurus (Matsushita et al. 2007b)}
\end{minipage}
\end{tabular}
\end{figure}



\subsection{Optical depth of Fe ${\rm_{XV\hspace{-.1em}I\hspace{-.1em}I}}$ in the central region}

We calculate the optical depth of resonance
scattering for the Fe ${\rm_{XV\hspace{-.1em}I\hspace{-.1em}I}}$ line 
in the central region mentioned in
$\S$3.4. We assume that the gas density distribution follows the $\beta$ model.
The scattering cross section $\sigma_\nu$ is represented by the 
following formula (Shigeyama et al. 1998),
\begin{eqnarray}
\sigma_\nu=\frac{\pi e^2}{m_e \nu_0} gf \sqrt[]{\mathstrut \frac{m_i}{2\pi kT}}
\end{eqnarray}
$m_i$ is the mass of the ion, $gf$ is oscillator strength, $\nu_0$ is
the frequency of the emission
and $T$ is the temperature of the system. We set $T$ to 0.642 keV which is
the temperature in the 0--2' region and the value of $gf$ to 2.5 
(Mewe et al. 1985).             
As a result, $\sigma_\nu$ becomes 1.20$\times$10$^{-15}$ cm$^2$.

For the density distribution $n(r)$, we use the $\beta$ model,
\begin{eqnarray}
n(r)=n_0[1+({\frac{r}{r_c})^2}]^{-{\frac{3}{2}}\beta}
\end{eqnarray} 
$n_0$ is the core density and $r_c$ the core radius.
For NGC 4636, the core density, the core radius and $\beta$ are
0.2 cm$^{-3}$, 1.1 kpc and 0.47, respectively (Ohto et al. 2003).

Then, the optical depth for the emission line from the circular region with
projected radius $R$ is calculated by the following formula,
assuming that the Fe abundance in the central region A$_{\rm Fe}$ is 1.14 solar
from the value summarized in Table 3, 
the Fe abundance solar ratio \rm{Fe/H} is
4.68$\times10^{-5}$ (Anders \& Grevesse 1989),  
ionic fraction $f_i$ of Fe ${\rm_{XV\hspace{-.1em}I\hspace{-.1em}I}}$ is 0.23 (SPEX) and
gas exists up to 300 kpc.
\begin{eqnarray}
\tau(r)=\int^{300}_R A_{\rm Fe} {\rm(Fe/H)} f_i \sigma_{\nu_0} n(r) \frac{r}{\sqrt[]{\mathstrut
 {r^2-R^2}}} dr
\end{eqnarray}
As a result of this calculation, the value of $\tau$ in the central
region within 2' becomes 24.1 $<$ $\tau$ $<$ 56.3.
Note that those values are not influenced by
thermal Doppler broadening and extent of the diffuse source.
Even if we consider them, $\tau$ $\gg$ 1 is secure.
Our result is somewhat larger than that of Werner et al. (2009), and the
difference is possibly explained by the assumption of the Fe abundance.
Therefore, it is optical thick there
and resonance scattering will easily occur.
Turbulence could reduce the resonance scattering (Werner et al. 2009),
but the scattering is significant for NGC 4636.

\section{Summary}

In this paper, we performed spectral analysis out to 28'($\simeq$ 140 kpc) from
the center of NGC 4636, 
using data from the XIS detectors onboard Suzaku. We obtained
the following results. 

\begin{itemize}
\item
{The temperature distribution shows a smooth gradient from
     the center (0.642 keV) to the outer regions. For the metal abundance,
     we found that the distributions of O, Mg, Si, Fe are all higher 
     in the central 4' region and decrease by 50\% toward the outer region.
     It is noticeable that O and Mg abundances are determined
     accurately, even out to 28' from the center.}
\item
{The abundance ratio of O/Fe is about 0.6$\sim$1.0 solar
     independent of the radius. This means that O and Fe in the gas 
     diffused to the outside of the galaxy after the products from both
     SNe II and SNe Ia were well mixed.}
\item
{We found out that the O and Fe radial mass distributions are
     similar between NGC 4636 and NGC 1399. 
     This indicates that both objects have
     similar metal diffusion processes. However the MLR of NGC 4636
     is 2$\sim$3 times larger than that of NGC 1399, possibly due to  
     NGC 4636 confining more metal-rich gas. The MLR of NGC 4636 is
     $\sim$5 times smaller than that of groups and clusters of
     galaxies and there may be a correlation between the temperature
     and MLR in other spherical groups.}

\item
{We confirmed resonance scattering in the 
     Fe ${\rm_{XV\hspace{-.1em}I\hspace{-.1em}I}}$ line in the
     central region of NGC 4636, first suggested by the
     XMM-Newton RGS results. 
     The optical depth in the central region is, 24.1 $<\tau<$ 56.3,
     and thus it is extremely optically thick in the central region if
     there is no turbulence.
    }
\end{itemize}

We thank to the anonymous referee for careful reading and many
helpful comments. We also thank to the Suzaku team for development of
hardware/software and operation.

 \end{document}